\documentclass[aps,pra,twocolumn]{revtex4}

\usepackage{color}
\usepackage[bottom]{draftcopy}   

\usepackage{graphicx}
\usepackage{sidecap}

\begin{document}

\title{Oblique-ECE Radial and Phase Detector of Rotating Magnetic Islands  
applied to Alignment and Phase-locked Modulation of ECCD for NTM 
Stabilization}


\author{F. Volpe$^{(a)}$}
\email[]{fvolpe@columbia.edu}
\author{M.E. Austin$^{(b)}$}
\author{G. Campbell$^{(c)}$}
\author{T. Deterly$^{(c)}$}
\affiliation{$^{(a)}$Columbia University, New York, NY, USA\break
$^{(b)}$University of Texas-Austin, Austin, TX, USA\break
$^{(c)}$General Atomics, San Diego, CA, USA}


\date{\today}

\begin{abstract}
A 2 channel oblique electron cyclotron emission (ECE) radiometer was 
installed on the DIII-D tokamak and interfaced to 4 gyrotrons. 
Oblique ECE was used to 
toroidally and radially localize rotating magnetic islands and so assist 
their electron cyclotron current drive (ECCD) stabilization. 
In particular, after manipulations operated by the interfacing analogue 
circuit, the oblique ECE signals directly modulated the current drive in 
synch with the island rotation and in phase with the island O-point, 
for a more efficient stabilization. 
Apart from the different toroidal location, 
the diagnostic view is identical to the ECCD launch direction,
which greatly simplified the real-time use of the signals. In fact, 
a simple toroidal extrapolation was sufficient to lock the modulation 
to the O-point phase. This was accomplished by a specially designed 
phase shifter of nearly flat response over the 1-7kHz range. 
Moreover, correlation
analysis of two channels slightly above and below the ECCD frequency allowed  
checking the radial alignment to the island, based on the fact that for
satisfactory alignment the two signals are out of phase. 
\end{abstract}


\maketitle


\section{Introduction} 
Neoclassical Tearing Mode (NTM) stabilization by  Electron Cyclotron
Current Drive (ECCD) requires good radial alignment  to the rotating
island, and is more efficient if the ECCD is modulated to
predominantly drive current in the  island O-point \cite{Rob_Rev}.
Alignment and modulation require diagnosing in real time  the radial
and toroidal position of the rotating island.  For alignment, this was
done by horizontal Electron Cyclotron  Emission (ECE) \cite{JT60-03,
  FTU}.  For modulation it was done by means of magnetic diagnostics
\cite{Marasch} and could in principle be done by horizontal ECE, which
is  spatially localized and can more accurately determine the toroidal
phase.  ECE was used extensively for this purpose as an off-line
diagnostic  \cite{TFTR-96, JT60-99, AUG-01, TEXTOR-03}.  However,
alignment and modulation assisted by magnetic probes and/or
horizontal ECE require real-time equilibria and helical extrapolations
from the measurement location to the  ECCD location,  taking into
account the 3D structure of the island \cite{Marasch, JT60-03}.  Apart
from being complicated, these extrapolations suffer from the
propagation  of systematic and statistical errors.



These complications become unnecessary if ECE is collected along the same 
{\em oblique} line of sight as ECCD \cite{Ooster} 
or an equivalent, toroidally displaced but otherwise identical 
line of sight \cite{PoP}. The present paper describes the DIII-D diagnostic 
setup for the latter approach, which led to the first oblique-ECE-assisted 
stabilization of NTMs \cite{PoP}. Emission was collected via a 
temporarily disconnected ECCD launcher, at nearly the same frequency and with 
the same view angles as ECCD and from the same radial and vertical location 
$R$ and $z$, but at a different toroidal location $\phi$. 
By reciprocity, this oblique emission detects the island at the same location 
where the EC current is driven, except for a toroidal -rather than helical- 
offset. This offset is easily corrected by appropriate electronics. 

Oblique ECE applied to ECCD is also immune from
systematic errors associated with refraction and diffraction, which are 
the same for the emitted and the injected millimeter (mm) wave beam. 
An exception to such symmetry between oblique ECE and ECCD is when the
third EC harmonic is present in the DIII-D plasma, causing 
measurable absorption and emission but negligible current drive. 

The radiometer (described in Sec.\ref{Sec_muWaveSetup} together with the 
front optics) features only 2 channels, for the reasons discussed in 
Sec.\ref{SecPrincip}. 
Here a single radiometer controls the modulation of 4 gyrotrons 
at different toroidal locations. This is done by means of a broad-band 
phase-shifter, and might have advantages over the in-line approach 
\cite{Ooster} in devices equipped with several gyrotrons, like ITER. 
Additionally, using distinct antennas and transmission lines for 
oblique ECE and ECCD reduces the technical difficulties of separating 
MW-level injected mm-waves from nW-level emission from the plasma. 
 
On the other hand, the introduction of the phase shifter comes with its own 
challenge: as the toroidal separation between launcher and receiver is fixed, 
the {\em phase}-delay needs to be fixed and frequency-independent. 
Most phase-shifters, however, apply a fixed {\em time}-delay. 
The problem was solved by a special $RC$ circuit 
where the resistance $R$ varies with frequency in a digitally controlled 
manner (Sec.\ref{SecInterf}). 

Finally, experimental results are presented in Sec.\ref{SecResults} and 
capabilities and limitations of the 
diagnostic technique are discussed in Sec.\ref{SecLimt}.

\section{Why only two channels}   \label{SecPrincip}
With a rough knowledge of the island radial location $R_{isl}$
relative to the emitting layer location $R_{ch}$, and whether they lie
on the low  or high field side (LFS, HFS), a single channel is 
sufficient to determine whether the island O-point is transiting in the 
oblique ECE view when the temperature fluctuation $\delta T_e$ is positive or 
negative (Table \ref{tab1}). 
Therefore, a single oblique ECE channel is sufficient to modulate the ECCD 
with the correct frequency and phase, with no ambiguity between O- and 
X-point. 

%

\begin{table}[b] 
\caption{\label{tab1} O-point discrimination from a single ECE
  channel (horizontal or oblique) measuring at radial location $R_{ch}$. 
  $R_{isl}$ denotes the island location. Obviously X-point detection criteria 
  are opposite to O-point criteria. For example, if the O-point is identified 
  by $\delta T_e >0$, the X-point is identified by $\delta T_e <0$, 
  and vice versa.}
\begin{ruledtabular}
\begin{tabular}{llc}
& Island Location & O-point \\
\hline
HFS  &  $R_{ch} < R_{isl}$  &  $\delta T_e >0$ \\
     &  $R_{ch} > R_{isl}$  &  $\delta T_e <0$ \\
LFS  &  $R_{ch} < R_{isl}$  &  $\delta T_e <0$ \\
     &  $R_{ch} > R_{isl}$  &  $\delta T_e >0$ 
\end{tabular}
\end{ruledtabular}
\end{table}

At least two channels, however, are necessary for radial alignment.
Their frequencies need to be above and below the ECCD frequency.  If
ECCD is correctly deposited at the island center, the two oblique ECE
signals will be out of phase with each other. If instead they are in
phase, they are originating from the same side of the island. This implies that 
ECCD is also being deposited on that side of the island rather than at its 
center, where it would be desirable.  

To clarify, let us consider ECCD on the HFS, well-aligned with the 
island of interest. The island is  ``bracketed'' between two channels 1 and 2
($R_1<R_{isl}<R_2$). Let the raw, uncalibrated signals 
$V_1$ and $V_2$ be non-inverted and proportional to $T_e$ 
at locations $R_1$ and $R_2$.  
The waveform $V_1$-$V_2$ will peak in correspondence of the O-point transit. 
Hence, it can be used to modulate the ECCD in synch and phase with the O-point. 
On the LFS, $V_2$-$V_1$ should be used. 

Note that, due to oblique view, emission is toroidally displaced with
respect to the receiver.  Similarly, due to oblique launch,
absorption is toroidally  displaced with respect to the launcher. The
two effects cancel each other, hence it is not necessary to correct
$V_2$-$V_1$ (or $V_1$-$V_2$)  for the emission-receiver or
launcher-absorption toroidal displacements.  However, it is necessary
to correct it for  the launcher-receiver toroidal phase-difference. 

Note also that for good radial alignment $| V_1-V_2 | \approx 2|V_1|$ 
and for poor radial alignment $| V_1-V_2 | \ll |V_1|$. Consequently, it is 
possible to chose a discrimination level (Sec.\ref{SecInterf}) below which 
$V_1-V_2$ is approximated with 0 at all times. In this way, no ECCD is injected 
when the alignment is wrong. 

This is simple but requires manual adjustment of the (absolute) 
discrimination level. 
For a more versatile discrimination, one can compare the sum and difference 
of the signals: 
for poor alignment $| V_1-V_2 | \ll | V_1+V_2|$, hence poor alignment can be 
defined by $| V_1-V_2 | / | V_1+V_2|$ falling below a relative threshold 
level.

\section{Experimental Setup}  \label{SecExpSet}
Fig.\ref{FigTopView} shows the top view of the experimental setup. 
As an NTM of toroidal mode number $n$ rotates with angular frequency 
$\omega$, it modulates -at frequency $n\omega/2\pi$- the 2nd harmonic 
X-mode emission originating at the dashed curve in Fig.\ref{FigTopView}. 
On the basis of this modulated emission, measured by a radiometer 
(Sec.\ref{Sec_muWaveSetup}), a circuit modulates the gyrotron power supplies 
with the correct frequency and phase-shift  
(Sec.\ref{SecInterf}) to drive EC current in the island O-point. 

\subsection{Millimeter Wave Setup}  \label{Sec_muWaveSetup}
A 2 channel heterodyne radiometer was built and installed at
DIII-D on the ECCD transmission line 5. 
The line of sight was set identical to the launch directions for 
gyrotrons 2, 3, 4 and 6, ensuring symmetry between 
emission and absorption. 
An optical switch allows to use launcher and transmission line 5  
for ECCD or oblique ECE (Fig.\ref{MicrowSetupFig}).
The optical switch has two sides at machine vacuum (a ``gyrotron'' 
side and a ``plasma'' side). A third side (the ``diagnostic'' side) 
consists of a short, evacuated corrugated overmoded waveguide with a quartz 
window. On the other side of the window, at atmospheric pressure, 
is the radiometer. A Teflon lens refocuses the divergent beam exiting from the 
waveguide window into a horn. 

Two additional quasi-optical components between the window and the horn select
the desired elliptical polarization and frequency. 
The first component is a metal plate array or ``artificial dielectric'' 
(Fig.\ref{FigQO}a). Electromagnetic waves of electric field
perpendicular to the plates propagate undisturbed. Waves of field parallel to
the plates, instead, propagate at higher phase velocity, $c/N$, where 
$N=\sqrt{1-(\lambda/2a)^2}$, $\lambda $=2.73mm is the wavelength and 
$a$=5.7mm the spacing between the plates. 
At the output of the 23.5mm long device,  
linear polarizations recombine with a shift of $\lambda/4$. 
Thus, linear polarizations not parallel nor perpendicular to the plates become
elliptical and, conversely, elliptical polarizations can be linearized. 
For oblique view, the O- and X-mode polarizations are, indeed, elliptical.  
Here the $\lambda$/4 shifter linearizes them and couples the X-mode to the
linearly polarized mm-wave components. This guarantees modal purity
and prevents mixing the main 2nd harmonic 
X-mode signal with the weaker (optically thin) but not negligible 
2nd harmonic O-mode (Fig.\ref{FigQO}).  
This would be undesirable because, as it 
experiences different refraction, it originates elsewhere, therefore the 
$\delta T_e$ information that it carries does not come from the intended 
location. 

To manufacture the device, we equally spaced and fixed in frame various 
double-sided copper clad glass-fiber boards of the type used in the
electronics industry. The thickness of the plates, 0.44mm, is much smaller
than their spacing $a$=5.7mm. At the same time, the 17$\mu m$ copper 
coating is much thicker than the 0.02$\mu m$ skin depth for 110GHz in 
copper, making the plates good reflectors. 

The other quasi-optical component is a dichroic plate (Fig.\ref{FigQO}b), 
basically a perforated Aluminum plate stopping waves of
wavelength bigger than the hole diameter. This is because, with a 95.75GHz 
local oscillator (LO) 
and 1GHz-wide channels centered at 12.5 and 15.5GHz, the radiometer is
not just sensitive to 107.75-111.75GHz, but also to the undesired 
79.75-83.75GHz band. This originates from the plasma edge and scrape-off
layer on the LFS.

For protection against stray radiation from EC heating and ECCD,
the radiometer is equipped with a stripline notch filter with rejection 
$>$40dB in the 110.05$\pm$0.45 GHz range, similar to the one
succesfully used in the horizontal ECE radiometer \cite{HorizRadiom}. 

The notch-filtered signal beats in a mixer with a Gunn diode LO,  
stabilized by a regulator, of measured frequency 
95.75GHz. 

The signal, now down-converted in frequency, is amplified by 
two GaAs Field Effect Transistor (FET) amplifiers of gain
33 and 25dB and low noise figures (2 and 7dB, respectively). 
Then it is split by a power divider and filtered by two bandpass filters 
centered at 12.5 and 15.5GHz. In fact, two sets of filters are available, 
respectively 400MHz and 1GHz wide.
The filters, in connection with the LO frequency, determine the channel 
frequencies of 108.25 and 111.25GHz, approximately symmetric around the gyrotron
frequency, 110GHz. Their separation, 3GHz, corresponds to a distance 
$\Delta R$=5cm in units of major radius. Note, however, that the island is 
viewed obliquely. 

The intermediate-frequency signals at the output of the filters
are measured by crystal detectors.  Their video-frequency outputs are
filtered ($f$<30kHz) and finally 
amplified by videoamplifiers of gain 
$G$=450-2300 and corresponding bandwidth 80-50kHz at 3dB, 
which is more than sufficient to assist ECCD up to the maximum 
gyrotron modulation frequency (10kHz).

\subsection{Interface between Radiometer and Gyrotrons, and Phase Shifter} 
\label{SecInterf}
As motivated in Sec.\ref{SecPrincip} and illustrated in 
Fig.\ref{FigBlockElectr}, a simple analog voltage subtractor generates 
the difference between the signals from the radiometer channels. 
The difference signal is then filtered to remove slow changes of $T_e$, 
noise, and NTMs which are too fast for gyrotron modulation, it is 
fed into the phase-shifter described below, and ultimately in a 
comparator. This converts it in a square wave suitable for modulation of the 
gyrotron power supply, but only if the input signal exceeds a threshold: small 
differential signals are not used for modulation, as they typically indicate 
poor alignment (Sec.\ref{SecPrincip}). 

The gyrotrons are located 15$^{\rm o}$ and 30$^{\rm o}$ away, 
toroidally, from the oblique ECE (Fig.\ref{FigTopView}).  
The desired phase shift is the sum of three contributions. 
The first one compensates for the different toroidal location of the 
oblique ECE sensor and ECCD actuators. Depending on the mode rotating  
clockwise or counter-clockwise, it can be a {\em phase} retardation  
($n15^{\rm o}$ or $n30^{\rm o}$, where $n$ is  the toroidal mode number)
or anticipation ($\Delta \phi =-n15^{\rm o}$ or -$n30^{\rm o}$). 
The second contribution is a fixed 
{\em time} anticipation, to compensate for the finite turn-on
and turn-off time of the gyrotrons and their power supply, as well as
the discharge time of their cables. It amounts to $\Delta t
_{gyr} =40$-$80\mu$s,  depending on the gyrotron. 
The third contribution, $\Delta t_f$, compensates for the  
fixed time-delays introduced by filters and videoamplifiers.  

The resultant of these contributions is always negative for
normal (counter-clockwise) plasma rotation. For clockwise rotation, it  
becomes negative at high frequencies.  
In the impossibility of imparting negative phase shifts, 
the circuit was designed for positive time shifts  
\begin{equation}
\Delta t = 
\frac{\Delta \phi}{360^o} \frac{2\pi}{\omega} -\Delta t_{gyr} -\Delta t_f
+N\frac{2\pi}{\omega},   \label{EqShift}
\end{equation}

where $N$ is the smallest integer giving $\Delta t>$0. 
As an example, a phase shift of -15$^{\rm o}$ is replaced by an 
equivalent phase shift of 345$^{\rm o}$. $N$ has to be small because the
frequency might change from one period to another. 

It is well-known that, generally, phase shifters provide constant time-delay 
$\Delta t$. 
Here, instead, we need a nearly-constant phase-shift $\Delta \phi$ 
(Eq.\ref{EqShift}). 
The range of interest is $f$=1-10kHz:  
modes rotating at less than 1kHz tend to lock, and other control 
strategies need to be used \cite{PoP}, while 10kHz exceeds the gyrotron 
modulation capabilities, in which case continuous ECCD can be used for 
stabilization. 

Constant phase shifts were obtained in the past 
in simple op-amp circuits with a fixed capacitor and a frequency-dependent 
resistor \cite{Japan,CERN}: with a resistance $R$ decaying inversely
with the frequency $f$, the $RC$-time of the circuit also decreases like
$1/f$, yielding costant phase shift. 
In turn, the frequency-dependent resistor consisted of a frequency-to-voltage
converter (FVC), of output $V \propto f$, 
and a voltage controlled resistor (VCR), typically a 
dual junction-gate FET (JFET), where the gate-source
voltage controls the resistance between the drain and the source.

Unfortunately this dependence is linear only in narrow ranges 
of frequencies: 20-40kHz in Ref.\cite{Japan} or 30-50MHz in Ref.\cite{CERN}. 
On the broader range of interest here (1-10kHz), 
dual JFETs and other FETs exhibit marked nonlinearities. 
For this reason 
it was decided to adopt as VCR a digitally programmable resistor 
controlled by a microcontroller (Fig.\ref{FigBlockPS}).  
The microcontroller has a built-in analog-to-digital converter
sampling the FVC output. 
In brief, the FVC measures the rotation frequency $f$ and 
converts in a voltage $V$ which is digitized by the microncontroller. 
On the basis of that input voltage $V$ and of a programmed $R(f)$ (or 
$R(V)$) curve, the microcontroller voltage-controls the VCR so that it 
takes the desired value of $R$. 
This is done virtually continuously, at a clock frequency of 20MHz, much
higher than 1-10kHz. 
The digital resistor is calibrated as follows. 
The phase-shifter is fed with sine-waves of known frequencies,  
then a variable resistor is adjusted so that the output
signal has the desired phase shift. The frequencies and associated resistor
values are recorded and a curve is fit to them.  
This curve is then programmed into the microcontroller.  

This approach extends the FVC-VCR technique to a
broader frequency range and adds flexibility, in that arbitrary curves 
$R(f)$ can be programmed, allowing for example the inclusion of   
$\Delta t_{gyr}$ and $\Delta t_f$ in Eq.\ref{EqShift}. 

The final result of this circuitry is a 0-10V square wave, optically converted 
and fed in the gyrotron power supplies. 
Four of these systems are implemented, with up to 7 interchangeable
microcontrollers, programmed for phase shifts equivalent to 0, $\pm 15$, $\pm
30$ and $\pm 60^{\rm o}$. The large shifts are intended for $n$=2 modes. 

Measurements confirm that the shifter introduces a dramatically non-flat 
$\Delta t$ and a nearly flat $\Delta \phi$ (Fig.\ref{FigProgRes}). 
The slight decrease of $\Delta \phi$ is intentional, because $\Delta t_{gyr}$ 
and $\Delta t_f$ become increasingly important at higher 
$f$. The good behavior of $\Delta \phi$ is the result of 
programming the microcontroller according to the manual calibration of $R(f)$  
described above. 

The shifter operates in the 1-7kHz range (Fig.\ref{FigProgRes}),
limited by the dynamic range and precision of the VCR, which  can only
be set to 256 values (8 bit) in the 0-10k$\Omega$  range. These
resistances are too low at the high end and  relatively too coarse at
the low end.  Extension to higher $f$ might be possible by   a
piecewise combination of curves of the type plotted in
Fig.\ref{FigProgRes},  adopting higher $N$ at higher $f$ (the digital
resistor can be  programmed to yield this complicated $R(f)$).   For
example, at $f>$7kHz one might apply $\Delta \phi$=705$^{\rm o}$
instead of 345$^{\rm o}$. Both are equivalent to -15$^{\rm o}$, but
705$^{\rm o}$ resets $R$ to a higher, more accessible value, from
which it  decays again with $f$, until eventually $N=3$ will be
adopted, and so on.

\section{Experimental Results}                             \label{SecResults}
The setup described here was used to 
control the ECCD modulation in real time in NTM stabilization 
experiments \cite{PoP}, as well as to check the quality of 
the radial alignment between ECCD and NTM after the discharge was completed. 

Fig.\ref{Fig132113det}a shows a detail of the ECCD power injected in the 
plasma, which was modulated on the basis of the oblique ECE signals in 
Fig.\ref{Fig132113det}b-c. The satisfactory 
quality of modulation is qualitatively confirmed by how well the ECCD
correlates with the Mirnov probe measurement of the rotating NTM 
in Fig.\ref{Fig132113det}d, except for the phase, which is
different, as it should be due to the different locations. 

The phase difference between the two oblique ECE channels is plotted in 
Figure~\ref{FigAlign1}. This is about zero, except in the marked 
interval, when the normalized minor radius 
$\rho$ of the ECCD matches the $\rho$ of the NTM within 1\%. 
Note in Fig.\ref{Fig132113det} and in the marked interval in 
Fig.\ref{FigAlign1} that the phase-difference between oblique ECE 
signals, even in case of good alignment, is not exactly $\pi$. This is 
expected, and discussed in the next Section.

\section{Capabilities and Limitations of the Technique}       \label{SecLimt}
The application of oblique ECE to ECCD modulation does not 
require the radiometer to be calibrated, neither absolutely (in temperature) 
nor relatively (from channel to channel). 
The application to ECCD alignment does not need absolute calibration either, 
but it requires the two channels to have approximately the same responsivity, 
because good alignment is recognized by two signals being out of 
phase and having approximately equal amplitude. 

The modulation works even when the alignment indicator does not, 
as far as one channel works (Sec.\ref{SecPrincip}). 
By lowering the threshold in the 
discriminator (Sec.\ref{SecInterf}), modulation can work even if 
good alignment has not been achieved yet and the signals are in phase, nearly 
canceling each other. 

Likewise, the indicator of alignment can work even under conditions 
preventing proper modulation, such as the mode rotating too fast for 
gyrotron modulation ($f>$10kHz), but not for oblique ECE detection ($f<$30kHz). 

It should be pointed out that signals emitted on opposite sides of the island 
are not exactly out of phase by 180$^{\rm o}$, consistent with the oblique 
view and with flow shear \cite{PoP}.
The effect is more pronounced (to the point that signals appear in phase) 
if the view is tangent to the flux-surface of interest. 
This geometry inhibits both modulation and alignment, and should be avoided. 

Another potential issue, paradoxically, stems from the fact that the
NTM and  ECCD have the  same frequency. As a result, at that frequency
the oblique ECE will  measure a $\delta T_e$ associated with the NTM,
and a $\delta T_e$ caused  by the modulated heating that inevitably
accompanies the modulated ECCD.  
In principle, this can lead to a nonlinear effect: 
the oblique ECE modulates the ECCD, which perturbs the oblique ECE, etc.
In practice, however, the radial profiles of $\delta T_e$ are so different 
that a phase discontinuity across the island center 
is recognized even in presence of a modulated ECCD background:  
the $\delta T_e$ associated with an NTM peaks at the island edges 
\cite{Rob_Rev} and  
$T_e$ basically oscillates as a whole, i.e.~with the same phase everywhere, 
except for a change of sign from one side of the island to the other; 
$\delta T_e$ from modulated ECCD, instead, peaks at the  
deposition location (i.e.~at the island center, if it is properly aligned) 
and propagates from there (heat pulses). 

Finally, an intrinsic limitation of oblique ECE is the higher level of 
Doppler broadening compared with horizontal ECE. 
This smooths and broadens the $\delta T_e(R)$ profile, and lowers its peaks, 
but preserves its shape and inversion radius, guaranteeing that modulation 
and alignment are still possible.

\begin{acknowledgments}
This work was supported in part by the US DOE under 
DE-FG03-97ER544156 and DE-SC0006415. 
FV thanks D.\ Kellman and R.\ Nazikian for the fruitful discussions and 
R.\ La Haye and R.\ Prater for the interest and encouragement. 
This paper is dedicated to coauthor T.\ Deterly, recently 
deceased. 
\end{acknowledgments}


\clearpage
\begin{widetext}
\begin{center}
  \begin{figure}[t]
       \includegraphics[scale=0.9]{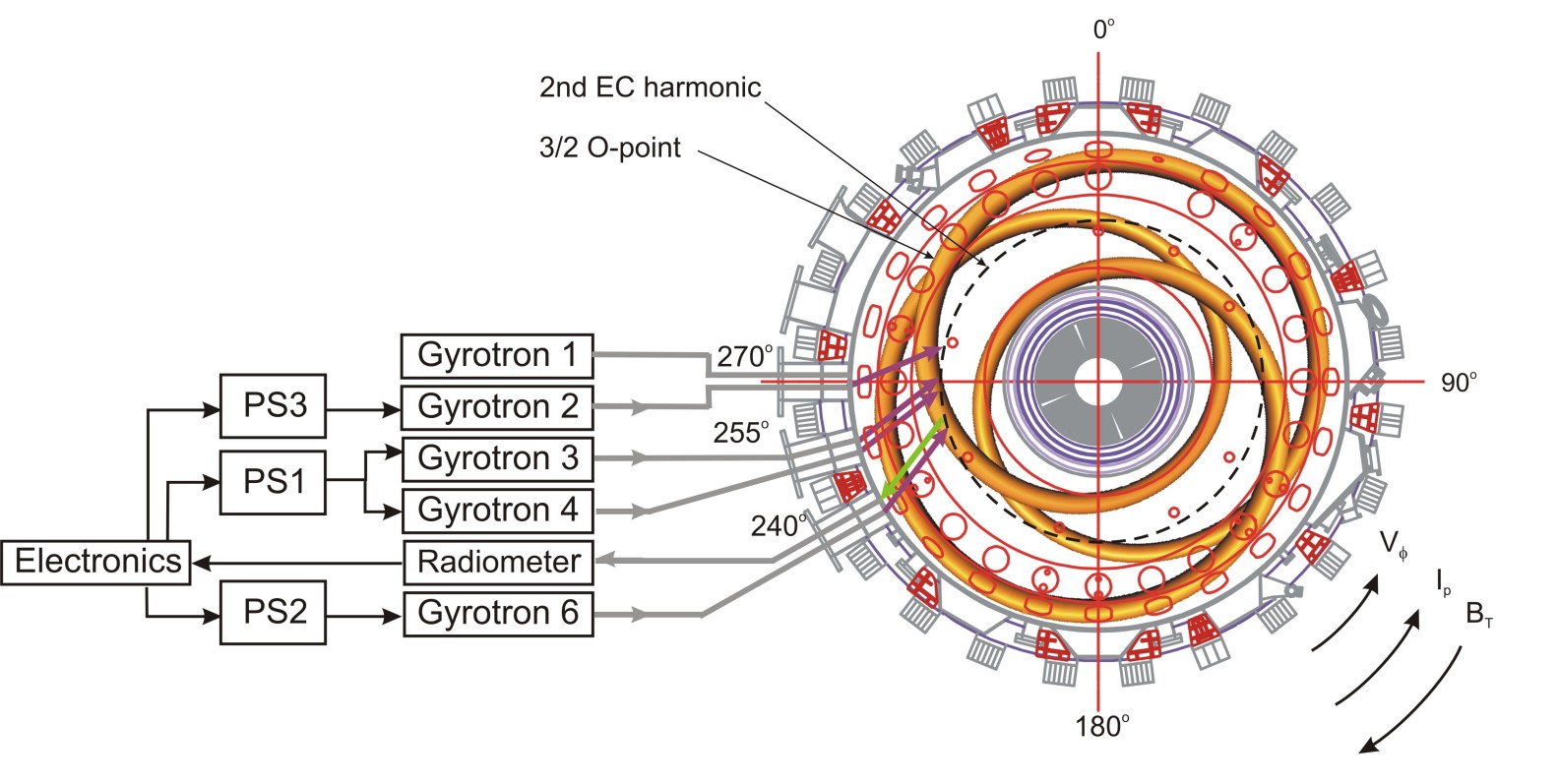}  
       \caption{\label{FigTopView} (Color online) 
       Top view of DIII-D and experimental setup interfacing 
       the oblique ECE radiometer to the gyrotron power supplies (PS), after 
       introducing a phase shift to compensate for the different toroidal 
       locations. 
       The radiometer collects emission from the dashed line, which varies 
       with time as the 3/2 mode rotates in the $v_\phi$ direction. 
       Only the O-point of the mode (or, more precisely, a flux tube around it) 
       is shown for simplicity, in orange.}
  \end{figure}
\end{center}

\begin{center}
  \begin{figure}[t]
      \includegraphics[angle=270,scale=0.95]{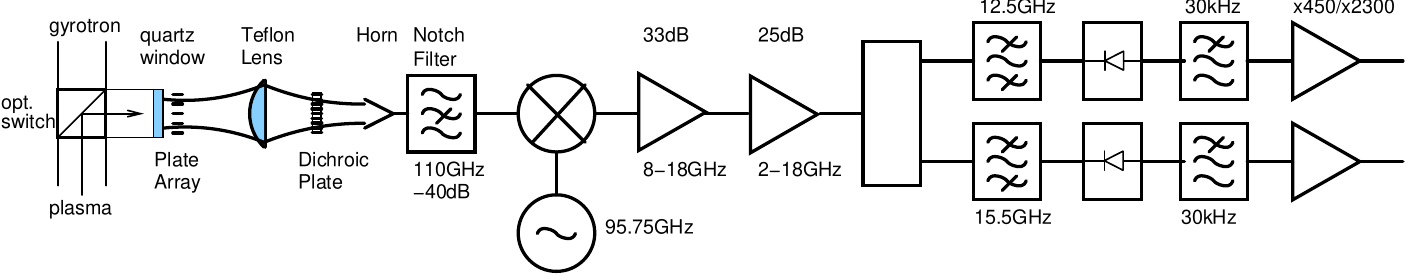} 
       \caption{\label{MicrowSetupFig} Schematic of the radiometer and 
         quasi-optics in front of it.}
  \end{figure}
\end{center}
\end{widetext}

  \begin{figure}[t]
       \includegraphics{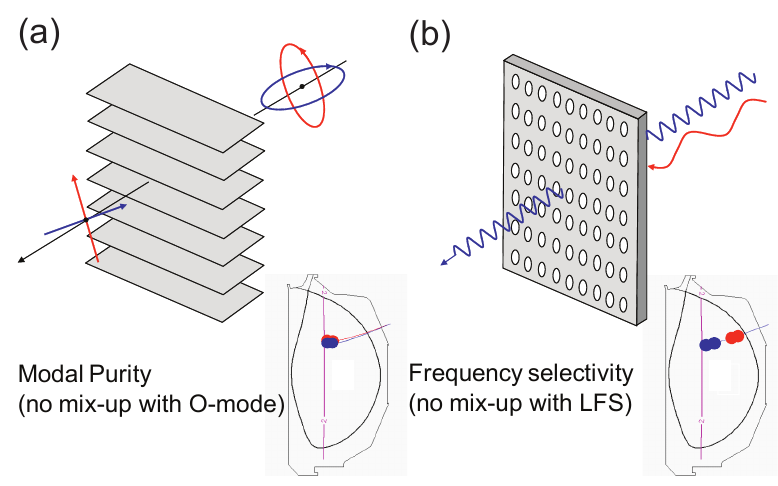} 
       \caption{\label{FigQO} 
         (Color online) (a) Plate array used to linearize the O and X 
         elliptical polarizations and guarantee pure X-mode measurements. 
         (b) Dichroic plate acting as a high pass filter, to reject the lower
         sideband (80-84GHz) of the radiometer. Insets: DIII-D cross-sections, 
         EC cold resonances and deposition regions for correct and incorrect
         polarizations and frequencies.}
  \end{figure}

\begin{figure}[t]
  \includegraphics{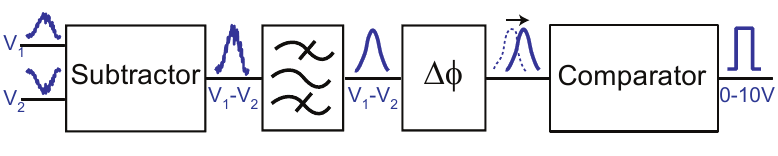} 
  \caption{\label{FigBlockElectr} 
    Block diagram of the electronics interfacing the oblique ECE radiometer 
    to the gyrotron power supplies, and schematics of how signals $V_1$ and 
    $V_2$ are combined and manipulated.}
\end{figure}

\begin{center}
  \begin{figure}[t]
    \includegraphics[scale=0.4]{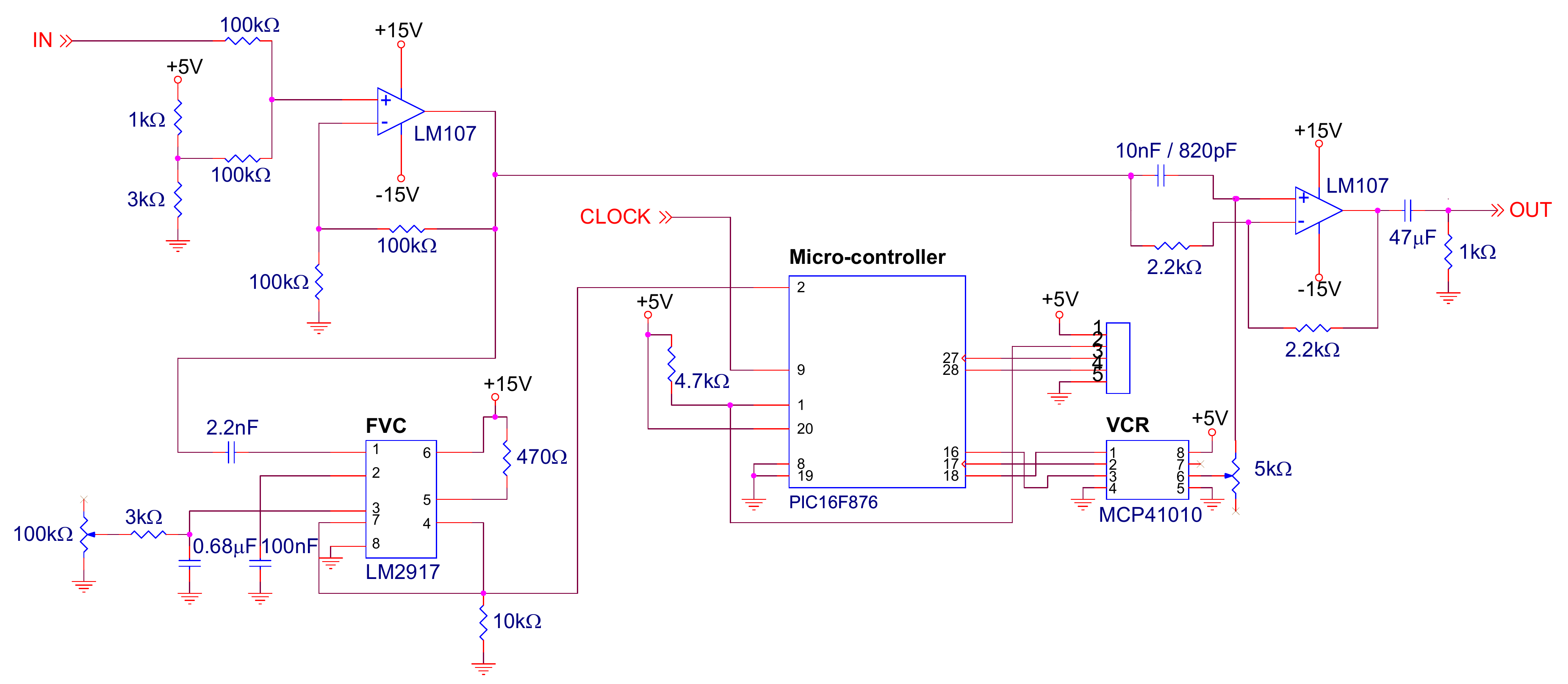} 
    \caption{\label{FigBlockPS} 
      Block diagram of the frequency-independent phase shifter. 
    A Frequency-to-Voltage Converter (FVC) controls, through a 
    micro-controller, a Voltage-Controlled Resistor (VCR). Its resistance $R$,  
    multiplied by the $C$=10nF/820pF capacitance, delays the 
    the signal by a time $RC$. The frequency-dependence of $R$ is programmed 
    to yield the phase-shift $\Delta \phi (f)$ of Eq.\ref{EqShift} and 
    Fig.\ref{FigProgRes}.}
  \end{figure}
\end{center}

\begin{figure}[t]
  \includegraphics[scale=0.85]{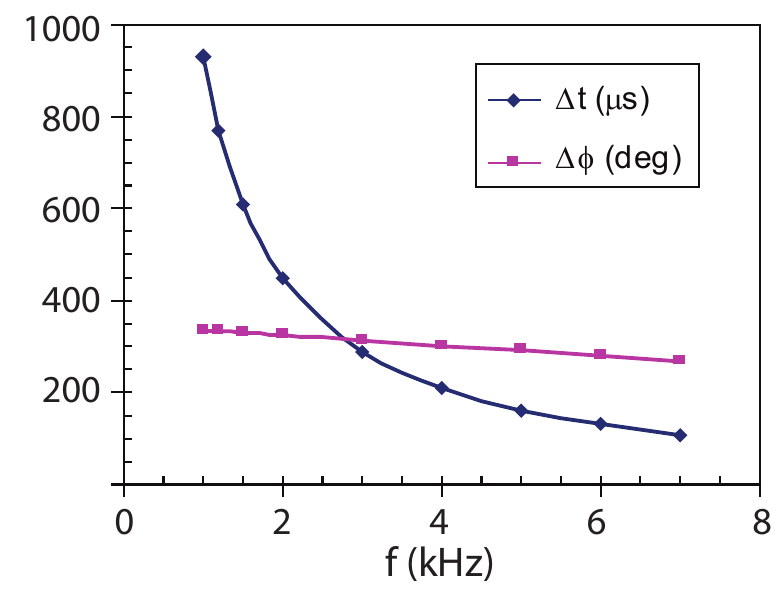} 
  \caption{\label{FigProgRes} (Color online) 
    Measured time-delay and phase-shift imparted by the circuit in 
    Fig.\ref{FigBlockPS} to gyrotrons 3 and 4, when the VCR is programmed for 
  a 3/2 mode rotating in the standard $v_\phi$ direction (Fig.\ref{FigTopView}).}
\end{figure}

\begin{figure}[t]
  \includegraphics[scale=1.]{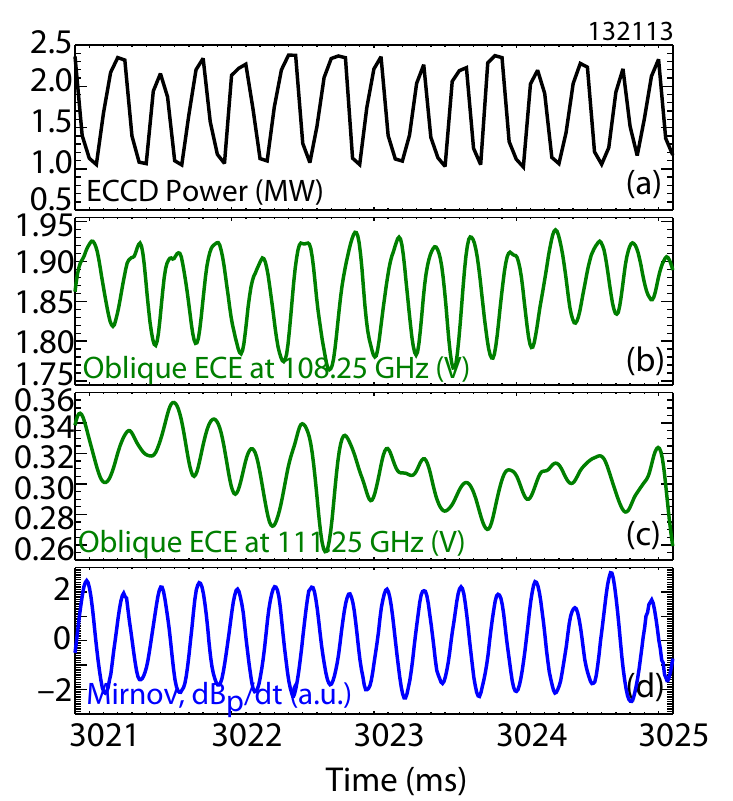} 
  \caption{\label{Fig132113det} (Color online) 
  (a)~Modulated ECCD in an NTM stabilization discharge \cite{PoP} 
generated by the difference between (b,c)~oblique ECE signals well-correlated
with (d)~Mirnov signal.}
\end{figure}

\begin{figure}[t]
  \includegraphics[scale=1.0]{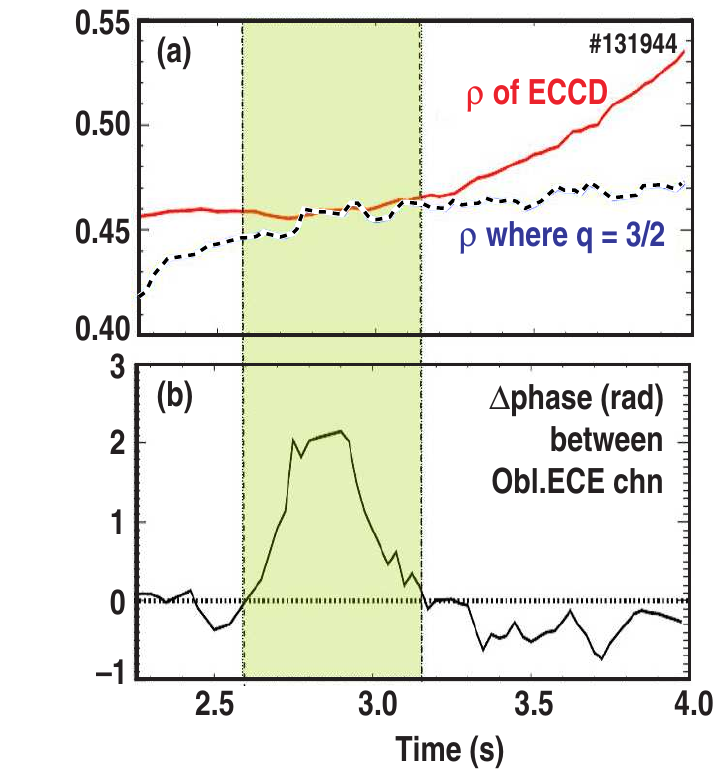} 
  \caption{\label{FigAlign1}(Color online)
 (a)~ECCD and $q=3/2$ normalized minor radii and (b) phase difference between
 oblique ECE signals at 108.25 and 111.25GHz, exhibiting a phase jump in 
 correspondence of good radial alignment between ECCD at 110GHz and the 
 $3/2$ rotating NTM responsible for the oblique ECE oscillations \cite{PoP}.}
\end{figure}

\end{document}